%% file: Population_Galactic_SNRs.tex
\documentclass[a4paper,11pt]{article}
\usepackage{pos}

\graphicspath{ {./Plots/} }
\usepackage{url}
\usepackage{hyperref}
\usepackage{natbib}
\usepackage{aas_macros}
\DeclareUnicodeCharacter{03B3}{$\gamma$}
\usepackage{caption}
\usepackage{subcaption}

\title{The population of Galactic supernova remnants
in the TeV range}
 \ShortTitle{The population of Galactic supernova remnants
in the TeV range}

\author*[a]{Rowan Batzofin}
\author[b]{Pierre Cristofari}
\author[a]{Kathrin Egberts}
\author[a]{Constantin Steppa}

\affiliation[a]{University of Potsdam,\\
  Am Neuen Palais 10, Potsdam, Germany}

\affiliation[b]{Observatoire de Paris, PSL Research University,\\
61 avenue de l’Observatoire, Paris, France}

\emailAdd{rowan.batzofin@uni-potsdam.de}
\emailAdd{pierre.cristofari@obspm.fr}
\emailAdd{kathrin.egberts@uni-potsdam.de}
\emailAdd{steppa@uni-potsdam.de}

\abstract{SNRs are likely to be significant sources of Galactic cosmic rays up to the knee. They produce gamma rays in the very-high-energy ($E>100$~GeV) range mainly via two mechanisms: hadronic interactions of accelerated protons with the interstellar medium and leptonic interactions of accelerated electrons with soft photons. Observations with current instruments have lead to the detection of about a dozen SNRs in VHE gamma rays and future instruments will help significantly increase this number. Yet, the details of particle acceleration at SNRs, and of the mechanisms producing VHE gamma-ray at SNRs remain poorly understood: What is the spectrum of accelerated particles? What is the efficiency of particle acceleration? Is the gamma-ray emission dominated by hadronic or leptonic origin?
\paragraph*{}
To address these questions, we simulate the population of SNRs in the gamma-ray domain, and confront it to the current population of TeV SNRs. This method allows us to investigate several crucial aspects of particle acceleration at SNRs, such as the level of magnetic field around SNR shocks or scanning the parameter space of the accelerated particles (spectral index, electron to proton ratio and the acceleration efficiency of the shock) with the possibility to constrain some of the parameters.
}

\ConferenceLogo{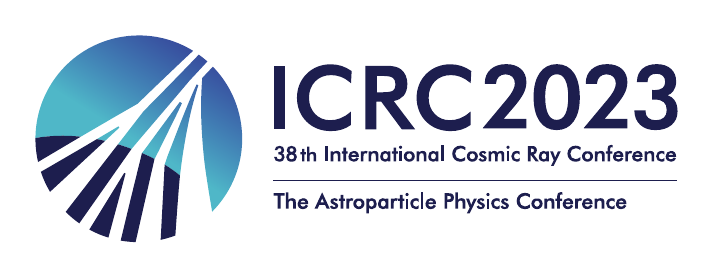}

\FullConference{%
38th International Cosmic Ray Conference (ICRC2023)\\
  26 July - 3 August, 2023\\
  Nagoya, Japan}

\begin{document}
\maketitle

\section{Introduction}
\label{section:intro}
Cosmic rays (CRs) were discovered thanks to the pioneer work of Domenico Pacini and Victor Hess in the 1910s and have been extensively studied ever since\citep{Hess:1912srp}. Remarkably, a fundamental question of CR physics remains unanswered: where are they produced? The standard paradigm for the origin of Galactic CRs is that supernova remnants (SNRs) accelerate the bulk of CRs, via the first-order Fermi mechanism called Diffusive shock acceleration (DSA)\citep{2010ApJ...718...31P}. DSA can efficiently energize ions (mostly protons), as well as electrons. Subsequently, the accelerated protons and electrons can interact with the interstellar medium (ISM) to produce gamma rays in the very-high-energy domain, mainly through two mechanisms: 1) the production/decay of neutral pions, in the collision of accelerated protons with hadrons of the ISM (hadronic mechanism), and 2) the inverse Compton scattering of accelerated electrons on soft photons (CMB, IR, optical) (leptonic mechanism). In most cases, it is very difficult to differentiate between a leptonic and a hadronic origin of the gamma-ray emission. The study of the population of SNRs in the VHE gamma-ray domain can help us understand particle acceleration, and discriminate between the two mechanisms. 
\paragraph*{}
In this work we investigate the properties of the population of VHE SNRs in the Milky Way. For this purpose we simulate populations of Galactic SNRs using a Monte Carlo method. We simulate the explosion times and positions of the supernovae and calculate the gamma-ray emission from the SNRs at the current time. We confront the simulations with measurements from the H.E.S.S. Galactic Plane Survey (HGPS) \citep{hgps_2018} by identifying the detectable sources in the simulations based on the longitude, latitude, and angular-extension dependent HGPS sensitivity and comparing it with the detected sample of SNRs in the HGPS. The inclusion of the complex dependencies of the HGPS sensitivity
is crucial for drawing conclusions on the SNR population based on this data comparison. Due to the large number of 47 unidentified sources in the HGPS the SNRs that are identified as such in the HGPS (8) are treated as a lower limit. Assuming that for at least some of the unidentified HGPS sources with SNR associations, their true nature is a SNR, we use the sum of actual SNRs and of HGPS sources associated (by position) with a known SNR (8 + 20) as the upper limit.  

\section{SNR population model} \label{section:pop_model}
The SNR population model is constructed from three basic ingredients: the physics of the SNR, the spatial source distribution, and the matter distribution. In this work we explore these ingredients to probe the properties of the Galactic population of SNRs.

\paragraph*{}
The SNR evolution follows the work presented in \cite{Cristofari_2013} and \cite{Cristofari_2021}. The dynamical evolution of the shock (radius and velocity as function of time) is computed following the approach of \cite{Chevalier_1982},\cite{Truelove_1999},\cite{Ptuskin_2003} and \cite{Ptuskin_2005}, taking into account the density of the interstellar matter \citep{Shibata_2011} at their positions. 
The magnetic field amplification, due to non resonant streaming is taken into account. We work under the assumption that a fraction $\xi_{CR}$ of the ram pressure is converted at the shock into CR protons; and that a fraction Kep of  $\xi_{CR}$ is converted into accelerated electrons.
The gamma-ray emission from the SNR is then calculated from the spectrum of protons and electrons, assuming an electron-proton ratio and a spectral index. 
In our approach, we are thus left with four main parameters accounting for particle acceleration: 1) the spectral index of the accelerated particles ($\alpha$), 2) the ratio of electrons to protons (Kep), 3) the acceleration efficiency ($\eta$) of the shock, and 4) the magnetic field amplification in the shock.

\begin{figure}[h]
\includegraphics[width = \textwidth]{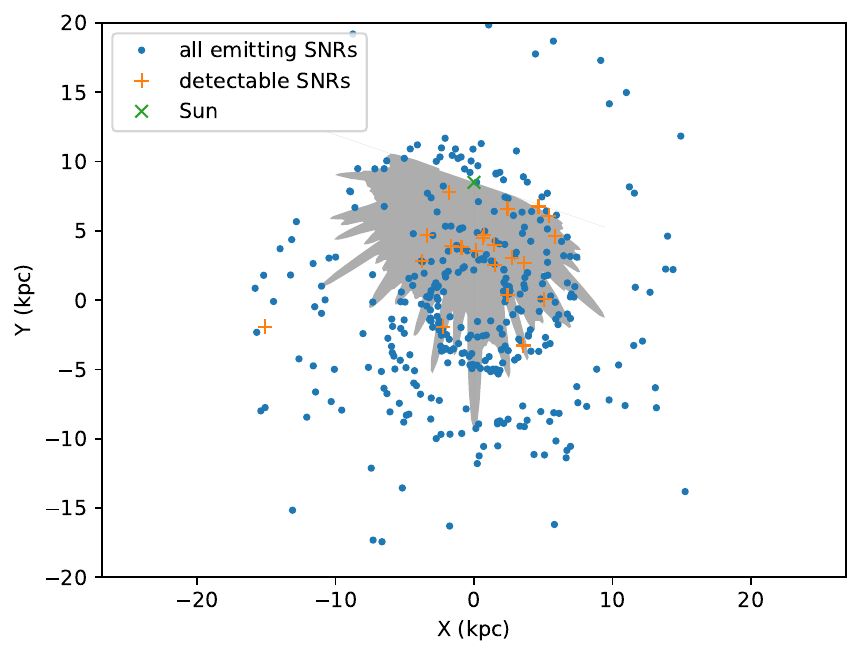}

\caption{Plot showing a single population of simulated SNRs. The gray shaded region is the HGPS detectability range for point like sources with a luminosity of $5\times 10^{33}$ photons~s$^{-1}$. The orange pluses are the simulated sources which are detectable and the blue dots are simulated SNRs which are not detectable. The location of our Sun is shown by a green cross.}
\label{fig:simulation_example}
\end{figure}

\paragraph*{}
 The SNR ages are drawn assuming a uniform distribution and assume a total supernova explosion rate of three per century. We use a Monte Carlo approach and distribute the SNRs in the Milky Way according to either a four-arm spiral model based on the interstellar medium \citep{Steiman-Cameron_2010} (referred to as Steiman, an example simulation is shown in Figure~\ref{fig:simulation_example}), a four-arm spiral model used to model pulsar positions \citep{CAFG_2006} (referred to as CAFG), a four-arm spiral arm model based on data from young massive stars with maser parallaxes \cite{Reid_2019} (referred to as Reid), or an azimuthally symmetric model with no spiral arms based on SNR data \citep{Green_2015} (referred to as Green). In this work we use a galactocentric coordinate system with the sun located at x = 0, y = 8.5 kpc, z = 0. The Steiman and Green models are both implemented following the work by \citep{Steppa_2020}. In order to account for the variety of SNRs, four SNR types are simulated, the relative rates and properties of each SNR type are summarized in Table \ref{table:types} \cite{Zirakashvili_2010},\cite{Cristofari_2013},\cite{CTA_Pevatron}. We consider that SNRs are efficiently accelerating particles up the to end of the   Sedov Taylor phase (typically of the order of 20 kyrs). 

\begin{table}[h]
\centering
\include{tables/type_table}
\caption{The four SNR types used in the simulations, with their properties and relative rates. $\epsilon_{51}$ is the ejecta energy in $10^{51}$ erg. $M_{ej,\odot}$ is the mass of ejecta in solar masses. The mass of ejecta is fixed for types Ia and IIb but normally distributed around around the mean displayed for IIP and Ib/c. $\dot{\text{M}}_{-5}$ is the wind mass loss rate in $M_{\odot}$/yr.}
\label{table:types}
\end{table}

\section{Confronting our simulated populations to available data} 
\label{section:simulating_populations}
We simulate 100 populations for each set of population properties. The positions of the sources are different for each of the 100 populations in a property set, however, if the source distribution property remains the same and other properties change those same 100 position sets are used. This allows us to closely examine how each of the properties affects the population.

\paragraph*{}
With the gamma-ray flux, size, and position we are able to determine whether each SNR would have been detected in the HGPS. To account for the selection bias in the H.E.S.S. catalog we follow the work in \citep{Steppa_2020} and in particular account for the fact that the HGPS sensitivity varies with position and source extent. In Figure \ref{fig:simulation_example} the HGPS detectability range for point like sources with a luminosity of $5\times 10^{33}$ photons~s$^{-1}$ is displayed, demonstrating the large variations in sensitivity as function of Galactic longitude.
\paragraph*{}
To test the different population properties we compare the integrated flux above 1 TeV of the simulated SNRs to that of the sources in the HGPS. Because many of the HGPS sources are unidentified and at least some of those should be SNRs we create a lower and upper limit for the comparison based on the number of identified SNRs (lower limit) and the number of identified or associated SNRs (upper limit).

\section{Results} \label{section:results} 
\begin{figure}[h]
\begin{subfigure}{0.5\textwidth}
\includegraphics[width = \textwidth]{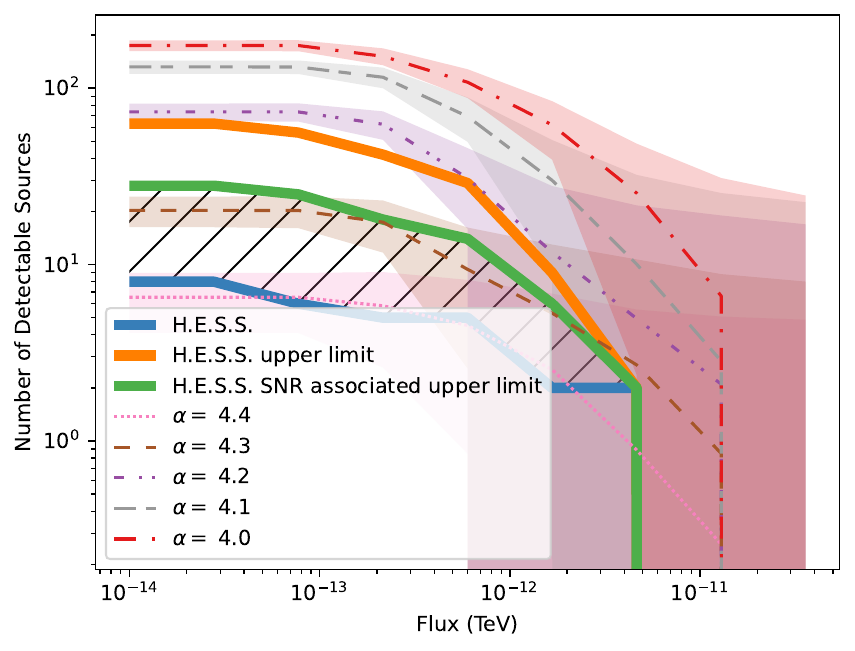}

\end{subfigure}
\begin{subfigure}{0.5\textwidth}
\includegraphics[width = \textwidth]{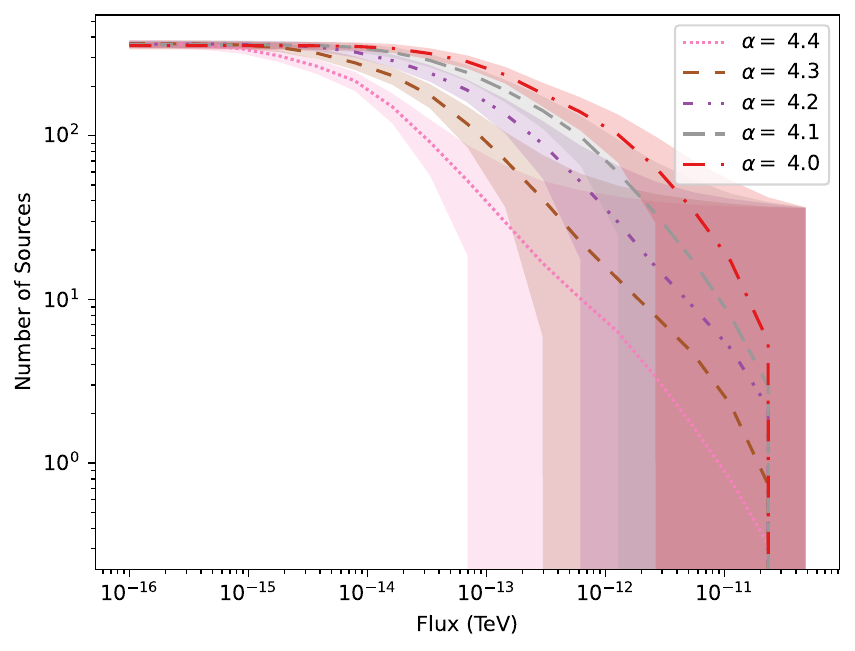}

\end{subfigure}
\caption{Cumulative distribution of the integrated flux above 1 TeV of the detectable simulated SNRs (left) and all simulated SNRs (right) averaged over the 100 populations with 1 standard deviation shown. The H.E.S.S. SNRs fluxes are shown in the thick blue line and the thick orange line shows the flux distribution for all the H.E.S.S. sources that could be SNRs i.e. not identified as something else. The thick green line shows the H.E.S.S. sources that have an association with a SNR, a more realistic upper limit. The hatched region shows the ideal area for the simulated SNRs to be in. The adjustable property of these populations is the spectral index of the accelerated particles, which is varied from 4 to 4.4. The other properties are kept constant: Kep = $10^{-4}$, $\eta = 6\%$, X-ray estimation is used for the magnetic field amplification, and the spatial distribution follows the Steiman model.}
\label{fig:flux_alpha_all}
\end{figure}

 Our simulations are confronted to the HGPS (figure \ref{fig:flux_alpha_all}). The cumulative distribution of the integrated flux of the sources (both simulated and H.E.S.S.). The spectral index is changed for each set of simulated populations and a decrease in spectral index increases the number of detectable sources in the population. Demonstrated in Figure \ref{fig:flux_alpha_all} (right) is that the total number of SNRs is unchanged.

\begin{figure}[h]
\includegraphics[width = 0.85 \textwidth]{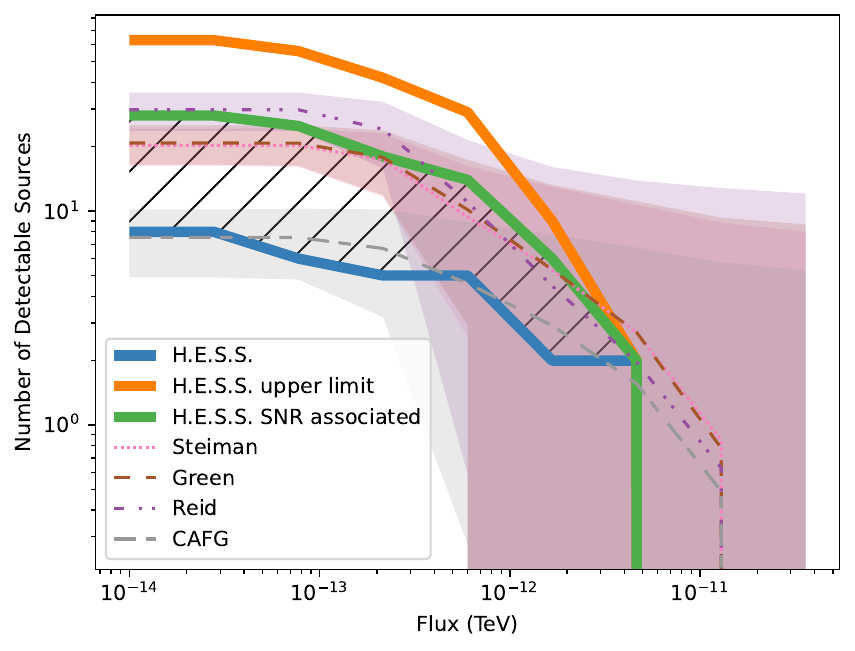}

\caption{Cumulative distribution of integrated flux of detectable simulated SNRs, the same as in Figure \ref{fig:flux_alpha_all} (left). The adjustable properties of these populations is the spatial distribution of the simulated SNRs. The constant properties are: $\alpha = 4.3$, Kep = $10^{-4}$, $\eta = 6\%$ and X-ray estimation for the magnetic field amplification method.}
\label{fig:flux_distribution}
\end{figure}

\begin{figure}[h]
\includegraphics[width = 0.85 \textwidth]{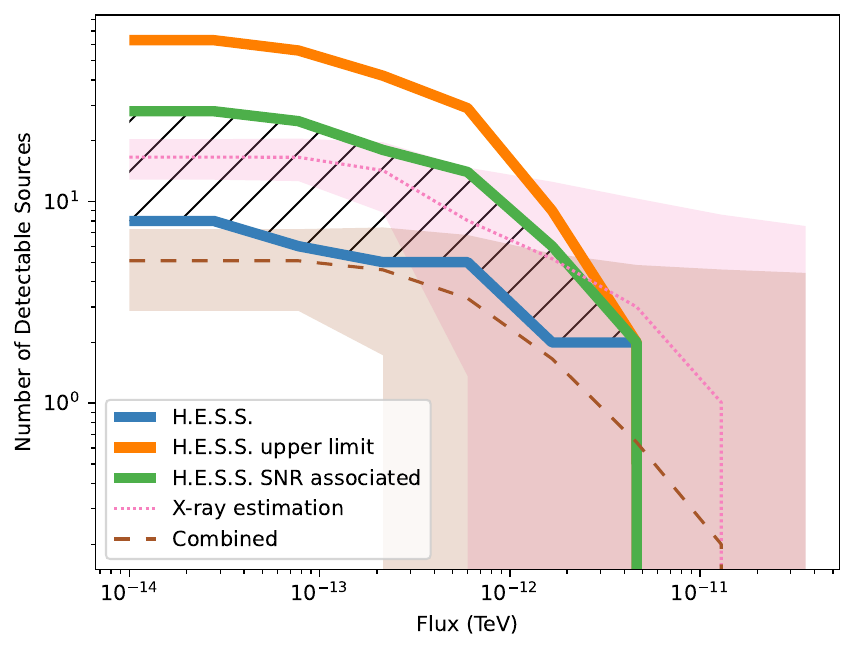}

\caption{Cumulative distribution of integrated flux of detectable simulated SNRs, the same as in Figure \ref{fig:flux_alpha_all} (left). The constant properties are: $\alpha = 4.3$, Kep = $10^{-4}$, $\eta = 6\%$ and the Steiman model as spatial distribution.}
\label{fig:magnetic_field}
\end{figure}

\paragraph{}
A variation of the spatial model is shown in Figure \ref{fig:flux_distribution}, clearly demonstrating that our results are not significantly impacted by changing from the Steiman to the Green distribution. In Figure \ref{fig:magnetic_field} we investigate the impact of the magnetic field amplification by considering two different methods: 
an estimation based on X-ray observations with the Hillas criterion \cite{Vokl_2005} and combining Bell amplification (<$\sim10$ kyr) \cite{Bell_2004} with resonant streaming \cite{Caprioli_2012} (10 kyr - 20 kyr).
\paragraph{}
Improvements in the processing speed of the simulations allow us to not only investigate the effect of varying a single property at a time, but also to explore the parameter space of $\alpha$, Kep and $\eta$ systematically. For the comparison with data, we use the percentage of the populations having a total number of detectable SNRs in the range between the number of H.E.S.S.-detected SNRs (the lower limit) and the H.E.S.S. sources associated with SNRs (the upper limit) as figure of merit. This range defined by the H.E.S.S. measurements is displayed as hatched region in Figures \ref{fig:flux_alpha_all}-\ref{fig:magnetic_field}. 

\begin{figure}[h]
\includegraphics[width = \textwidth]{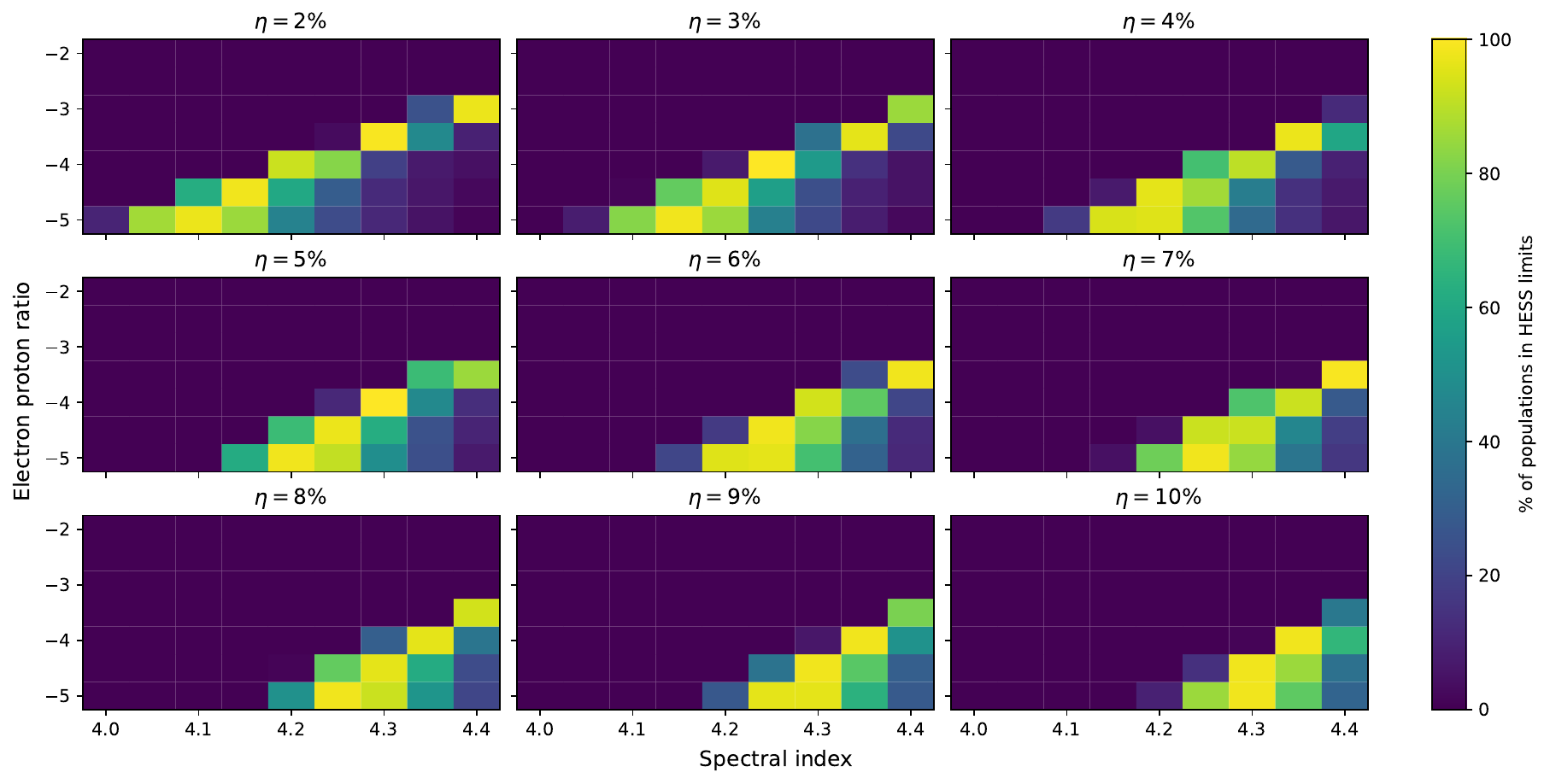}
\caption{Scan of the parameter space of the SNR model: shown is the percentage of populations that have at least as many detectable simulated SNRs as firmly detected H.E.S.S. SNRs and no more than the H.E.S.S. sources with associated SNRs for parameter combinations of the electron-proton ratio, the spectral index, and the acceleration efficiency. The electron-proton ratio is shown as logarithm. The Steiman spatial distribution and the magnetic field amplification estimation based on X-ray observations are used for all populations. Bright colours indicate a good agreement with data, for dark colours the parameter combination is in conflict with HGPS observations.}
\label{fig:eta_subplots}
\end{figure}

The typical ranges explored are: $\alpha$ from 4 to 4.4, Kep from $10^{-2}$ to $10^{-5}$ and $\eta$ from 1\% to 10\%.
\paragraph{}
The results of the parameter exploration are shown in Figure \ref{fig:eta_subplots}. We observe clear correlations between parameters: and increase in the cosmic-ray efficiency is compensated by a decrease in the electron-proton ratio, or an increase in the spectral index.
These intertwining make it arduous to identify a preferred region  of the parameter space, but it is still possible to exclude some part of the parameter space. The population is better explained by more hadronic sources and we can rule out populations with an electron to proton ratio of $> 10^{-3}$. The population is also better described by a steeper spectral index and rules out populations with $\alpha = 4$. The SNR population is best described by particles with a spectral index between 4.1 and 4.4 and an electron to proton ratio between $10^{-3}$ and $10^{-5}$. 

\section{Conclusion}
We present here a novel realistic comparison of models of the Galactic SNR population in the VHE range with the available TeV data, taking into account the multi-dimensional exposure of the HGPS. We especially report on an extensive exploration of the parameter space including the usual parameters describing particle acceleration at SNRs : the electron-proton ratio, the spectral index, the CR acceleration efficiency, the estimation of the magnetic field amplification. While correlations prevent the identification of an optimal combination of the evaluated parameters, the H.E.S.S. data have the potential to constrain parameter ranges for the chosen setup of simulations. Thus we find e.g. a tendency for lower electron-proton ratios, with values larger than $10^{-3}$ basically excluded and for steeper spectral indices.  

Finally,  the observed degeneracy could be alleviate by  any additional constraint on one of the discussed parameters, that could help substantially narrow the allowed parameter space. 
\bibliography{Population_Galactic_SNRs}

\end{document}

%% file: tables/type_table.tex
\begin{tabular}{|c|c|c|c|c|}
\hline
Type & Ia & IIP & Ib/c & IIb \\
\hline
Rel. rate & 0.32 & 0.44 & 0.22 & 0.02 \\
$\epsilon_{51}$ & 1 & 1 & 1 & 3 \\
$\text{M}_{ej,\odot}$ & 1.4 & 14 & 2 & 1 \\
$\dot{\text{M}}_{-5}$ & - & 1 & 1 & 10 \\
\hline

\end{tabular}

%% file: Population_Galactic_SNRs.bbl
\begin{thebibliography}{10}

\bibitem{Hess:1912srp}
Victor~F. Hess.
\newblock {\"Uber Beobachtungen der durchdringenden Strahlung bei sieben
  Freiballonfahrten}.
\newblock {\em Phys. Z.}, 13:1084--1091, 1912.

\bibitem{2010ApJ...718...31P}
Vladimir et~al. {Ptuskin}.
\newblock {Spectrum of Galactic Cosmic Rays Accelerated in Supernova Remnants}.
\newblock {\em \apj}, 718(1):31--36, July 2010.

\bibitem{hgps_2018}
H.E.S.S. Collaboration.
\newblock The h.e.s.s. galactic plane survey.
\newblock {\em A\&A}, 612:A1, 2018.

\bibitem{Cristofari_2013}
P.~Cristofari et~al.
\newblock Acceleration of cosmic rays and gamma-ray emission from supernova
  remnants in the galaxy.
\newblock {\em Monthly Notices of the Royal Astronomical Society},
  434(4):2748--2760, aug 2013.

\bibitem{Cristofari_2021}
{Cristofari, P.} et~al.
\newblock Cosmic ray protons and electrons from supernova remnants.
\newblock {\em A\&A}, 650:A62, 2021.

\bibitem{Chevalier_1982}
R.~A. {Chevalier}.
\newblock {Self-similar solutions for the interaction of stellar ejecta with an
  external medium.}
\newblock {\em \apj}, 258:790--797, July 1982.

\bibitem{Truelove_1999}
J.~Kelly {Truelove} and Christopher~F. {McKee}.
\newblock {Evolution of Nonradiative Supernova Remnants}.
\newblock {\em \apjs}, 120(2):299--326, February 1999.

\bibitem{Ptuskin_2003}
{Ptuskin, V. S.} and {Zirakashvili, V. N.}
\newblock Limits on diffusive shock acceleration in supernova remnants in the
  presence of cosmic-ray streaming instability and wave dissipation.
\newblock {\em A\&A}, 403(1):1--10, 2003.

\bibitem{Ptuskin_2005}
{Ptuskin, V. S.} and {Zirakashvili, V. N.}
\newblock On the spectrum of high-energy cosmic rays produced by supernova
  remnants in the presence of strong cosmic-ray streaming instability and wave
  dissipation.
\newblock {\em A\&A}, 429(3):755--765, 2005.

\bibitem{Shibata_2011}
T.~Shibata et~al.
\newblock A possible approach to three-dimensional cosmic-ray propagation in
  the galaxy. iv. electrons and electron-induced γ-rays.
\newblock {\em The Astrophysical Journal}, 727(1):38, dec 2010.

\bibitem{Steiman-Cameron_2010}
Thomas Y. Steiman-Cameron et~al.
\newblock Cobe and the galactic interstellar medium: Geometry of the spiral
  arms from fir cooling lines.
\newblock {\em The Astrophysical Journal}, 722(2):1460, sep 2010.

\bibitem{CAFG_2006}
Claude-André Faucher-Giguère and Victoria~M. Kaspi.
\newblock Birth and evolution of isolated radio pulsars.
\newblock {\em The Astrophysical Journal}, 643(1):332, may 2006.

\bibitem{Reid_2019}
{Reid, M.~J.} et~al.
\newblock {Trigonometric Parallaxes of High-mass Star-forming Regions: Our View
  of the Milky Way}.
\newblock {\em \apj}, 885(2):131, November 2019.

\bibitem{Green_2015}
D~A Green.
\newblock {Constraints on the distribution of supernova remnants with
  Galactocentric radius}.
\newblock {\em Monthly Notices of the Royal Astronomical Society},
  454(2):1517--1524, 10 2015.

\bibitem{Steppa_2020}
{Steppa, Constantin} and {Egberts, Kathrin}.
\newblock Modelling the galactic very-high-energy source population.
\newblock {\em A\&A}, 643:A137, 2020.

\bibitem{Zirakashvili_2010}
V.~N. {Zirakashvili} and F.~A. {Aharonian}.
\newblock {Nonthermal Radiation of Young Supernova Remnants: The Case of RX
  J1713.7-3946}.
\newblock {\em \apj}, 708(2):965--980, January 2010.

\bibitem{CTA_Pevatron}
CTA Collaboration.
\newblock Sensitivity of the cta to spectral signatures of hadronic pevatrons
  with application to galactic supernova remnants.
\newblock {\em Astroparticle Physics}, 150:102850, 2023.

\bibitem{Vokl_2005}
{V\"olk, H. J.} et~al.
\newblock Magnetic field amplification in tycho and other shell-type supernova
  remnants.
\newblock {\em A\&A}, 433(1):229--240, 2005.

\bibitem{Bell_2004}
A.~R. {Bell}.
\newblock {Turbulent amplification of magnetic field and diffusive shock
  acceleration of cosmic rays}.
\newblock {\em \mnras}, 353(2):550--558, September 2004.

\bibitem{Caprioli_2012}
Damiano {Caprioli}.
\newblock {Cosmic-ray acceleration in supernova remnants: non-linear theory
  revised}.
\newblock {\em \jcap}, 2012(7):038, July 2012.

\end{thebibliography}
